\documentclass[aps,prb,twocolumn,floats]{revtex4}

\usepackage{ifthen}
\usepackage{ifpdf}

\usepackage{latexsym}
\usepackage{amsmath}
\usepackage{amssymb}
\usepackage{bm}

\ifpdf
\usepackage{graphicx}
\usepackage{epstopdf}
\else
\usepackage{graphicx}
\usepackage{epsfig}
\fi

\newcommand{\trc}{\mbox{trace}}

\newcommand{\im}{\mbox{Im}}

\newcommand{\eexp}{\mbox{e}^}
\newcommand{\bra}{\left\langle}
\newcommand{\ket}{\right\rangle}

\newcommand{\tbox}[1]{\mbox{\tiny #1}}
 
\newcommand{\beq}[1]{\begin{eqnarray}\ifthenelse{#1=-1}{\nonumber}{\ifthenelse{#1=0}{}{\label{e#1}}}}
\newcommand{\eeq}{\end{eqnarray}} 

\newcommand{\be}{\begin{eqnarray}}
\newcommand{\ee}{\end{eqnarray}}

\newcommand{\Pcl}{{\tilde{P}}^{(0)}}
\newcommand{\bmq}{q}
\newcommand{\GammaSP}{\Gamma_{\varphi}}
\newcommand{\deltaF}{\delta_{\tbox{F}}}

\begin{document}

\title{The dephasing rate formula in the many body context}

\author{Doron Cohen$^1$, Jan von Delft$^2$, Florian Marquardt$^2$, and Yoseph Imry$^3$}

\affiliation{ $^1$\mbox{Department of Physics, Ben-Gurion University,Beer-Sheva 84105, Israel} \\ 
$^2$\mbox{Ludwig Maximilians Universitat Munchen, Theresienstr. 37, 80333 Munich, Germany} \\ 
$^3$\mbox{Department of Condensed Matter Physics, Weizmann Institute of Science, Rehovot 76100, Israel}}

\begin{abstract}
We suggest a straightforward approach to the calculation of the
dephasing rate in a fermionic system, which correctly keeps track of
the crucial physics of Pauli blocking. Starting from Fermi's golden
rule, the dephasing rate can be written as an integral over the
frequency transferred between system and environment, weighted by
their respective spectral densities. We show that treating the full
many-fermion system instead of a single particle automatically
enforces the Pauli principle. Furthermore, we explain the relation to
diagrammatics. Finally, we show how to treat the more involved
strong-coupling case when interactions appreciably modify the
spectra. This is relevant for the situation in disordered metals,
where screening is important.
\end{abstract}


\maketitle

\section{Introduction}
\label{sec:intro}

When a quantum system is coupled to an environment, one central
feature of the resulting dynamics is that the quantum system undergoes
dephasing, because its degrees of freedom get entangled with those of
the environment. Depending on context, a large variety of approaches
have been developed for calculating the dephasing rate. In the context
of the dephasing of electrons in disordered conductors, as measured,
e.g., via weak localization, these include (giving a partial list
only): a path integral [\onlinecite{AAK}] method to solve the
diffusion equation for the Cooperon; diagrammatic perturbation theory
[\onlinecite{Fukuyama1983,alt,alt2,Aleiner2002,munich2}]; Fermi golden
rule (FGR) arguments for the rate of energy exchange between the system
and the environment [\onlinecite{dph2,dph,imry,munich1}]; an approach relating
the loss of phase of propagating electrons to the change of state of
their environment [\onlinecite{SAI,imry}] or to the loss of purity [\onlinecite{rgd,rgd2}]; 
a semiclassical approach [\onlinecite{cak}] using the paradigm of particle plus (effective)
bath; and elaborations of this idea in terms of Feynman-Vernon-type
influence functionals for (quantum) Nyquist noise
[\onlinecite{zaikin,zaikin2,dph2,dph,imry,Delft2008,munich1}].

These methods vary greatly in their level of rigor and/or physical
transparency, and in the level of sophistication employed in dealing
with the subtleties that arise due to the indistinguishability of the
electron that is being dephased from the other electrons that dephase
it.  The associated "Pauli constraints" determine the fate of
dephasing in the low-temperature limit
[\onlinecite{alt,alt2,dph,dph2,Delft2008,munich1,munich2}]. 
ensuring that the dephasing rate vanishes in the zero 
temperature limit [\onlinecite{Delft2008}], 
contrary to some other claims [\onlinecite{zaikin,zaikin2}].

The present paper provides a pedagogical and 
physically transparent discussion of the role of the Pauli
constraints, without undue pretense of generality or rigor.  
We do this within the framework of the so-called SP-approach
[\onlinecite{dph2,dph,imry}]. Starting from the FGR, 
it expresses the dephasing rate as a ${d\omega dq}$ integral 
over a product  ${\tilde{S}(\omega, q) \tilde{P}(-\omega, -q)}$ 
that involves two unsymmetrized spectral functions~[\onlinecite{asym1,asym2}]. 
The first spectral function ($\tilde{S}$) describes the fluctuations of the environment,  
while the other ($\tilde{P}$) is the power spectrum of the density fluctuations 
of the system which characterizes the motion of the particles. 
The clear factorization of the relevant physics into system and bath 
that can exchange frequency ($\omega$)  and momenta ($q$)
is the main distinguishing feature of this approach.

Previous works using the SP-approach had employed a function $\tilde{P}^{[1]}$ 
which described the spectrum of a \emph{single} particle
propagating in a fermionic environment. To incorporate the physics of
Pauli blocking, which plays an essential role in determining the upper
cutoff on the frequency integral, a rather heuristic mix of
semiclassical and many-body arguments had been employed. The present
paper aims to rephrase the discussion of $\tilde{P}$ in the more
general context of an $N$-body system.  The corresponding spectrum
$\tilde{P}^{[N]}$ can be written down using standard and unambiguous
many-body constructions, without recourse to semiclassical arguments,
with Pauli-blocking factors arising in a very natural and standard
manner.  Remarkably, it turns out that $\tilde{P}^{[N]}$ is
proportional to $\tilde{P}^{[1]}$, with the proportionality factor 
${N_T = 2T/\deltaF}$ which is the effective number of thermally 
excited particles that can be scattered in a system that 
has a mean level spacing~$\deltaF$. This result justifies the way in which
Pauli-blocking factors had previously been built into $\tilde{P}^{[1]}$
by hand, and places it in a more general many-body context. It also
clarifies the relation of the SP-approach to the influence functional
approach [\onlinecite{Delft2008,munich1}] for dealing with a fermionic 
system under the influence of quantum Nyquist noise.

The structure of this paper is as follows.
Section~\ref{sec:quantumnoiseapproach} briefly reviews the very
general ``quantum noise approach'' to the problem of dephasing of a
quantum system with a discrete spectrum coupled to an environment.
Section~\ref{sec:qn-manybody} shows that when this approach is
generalized to a \emph{many-body} system interacting with an
environment, and the FGR is invoked to calculate the dephasing
rate, one readily arrives at an expression of the SP-type, with
$\tilde{S}$ and $\tilde{P}$ being unsymmetrized spectral functions for
system and environment. Section~\ref{sPqomega} discusses the
calculation of $\tilde{P}$ for \emph{noninteracting} fermions in various contexts: 
$\Pcl$ for a single particle; 
$\tilde{P}^{[e]}$ or $\tilde{P}^{[h]}$ for an electron or hole excitation in a Fermi sea;  
$\tilde{P}^{[1]}$ for the thermally averaged single-particle excitation of the Fermi sea; 
and $\tilde{P}^{[N]}$ for the entire many-body system.
In Section~\ref{sMetal} we use SP-theory to calculate the dephasing rate
of a many-body system \emph{weakly} coupled to an environment, and
establish the remarkable relation $\tilde{P}^{[N]} = (2 T/\deltaF)
\tilde{P}^{[1]}$ mentioned above. Finally, Sections~\ref{sDiagrm} and~\ref{sInteract} 
discuss the case that the system and the environment are so strongly coupled
that screening takes place, which modifies the system and/or particle
spectral functions, and puts it into perspective relative to
diagrammatic approaches. Several appendices summarize some
technical details, including the relation of the SP-approach to
a purity-based definition of dephasing recently introduced 
in [\onlinecite{rgd,rgd2}].

\section{The quantum noise approach and dephasing for simple systems}
\label{sec:quantumnoiseapproach}

When considering a quantum system coupled to a dissipative environment,
it is useful to apply the perspective of what we term the "quantum noise approach".
This means that, at weak coupling, all the effects of the environment on
the system (dissipation, heating, and dephasing) can be described completely
once the frequency spectrum of the noisy force coupling to the system is known.
For a recent general review of the quantum noise approach, especially in the context of quantum measurement and amplification, see [\onlinecite{2008_10_ClerkEtAl_QuantumNoiseReview}]. For our specific application to dephasing, we will later employ the scheme described in [\onlinecite{dph,dph2}].

In the present section, we will briefly review these ideas for the case of a general quantum system with a discrete spectrum (e.g. a two-level system). In the following section, we will then extend these considerations, both by keeping track of the spatial degree of freedom and by preparing the ground for a treatment of many-body systems, which is the emphasis of this article.

Consider a quantum system that couples to some fluctuating 
force field ${\hat{\mathcal{F}}}$, such that the interaction is
\be
\hat{V} \ = \ \hat{\mathcal{A}} \hat{\mathcal{F}} \, . \label{simpleV}
\ee
The FGR is an expression for the transition rate 
from an eigenstate~$n$ to some other eigenstate~$m$ 
of the system, and it can be written as:
\be
\Gamma_{m \leftarrow n} = \left|\left\langle  m \left| \hat{\mathcal{A}} \right| n \right\rangle \right|^2 
\times {\tilde{S}}(\omega = (E_m{-}E_n)), 
\label{FGRsimple}
\ee
Here the quantum noise spectrum ${\tilde{S}}(\omega)$ 
is defined as the Fourier transform (FT) 
of the autocorrelation function,
\be
{\tilde{S}}(\omega) 
= \mbox{FT}\left[ \left\langle  \hat{\mathcal{F}}(t)\hat{\mathcal{F}}(0)\right\rangle\right] 
= \int \left\langle  \hat{\mathcal{F}}(t)\hat{\mathcal{F}}(0)\right\rangle  e^{i\omega t} dt \, . \label{QNspectrumsimple}
\ee
Crucially, such a quantum noise spectrum is asymmetric, in contrast to the case of a classical stochastic process. This asymmetry contains important physics: The weight 
of the spectrum at positive frequencies indicates the rate of processes where energy is released into the environment, whereas the weight at negative frequencies belongs to those transitions where the system receives some energy.

It is now a straightforward observation that the {\em total} decay rate out of some given level can be simplified by introducing the spectrum ${\tilde{P}}(\omega)$ of the system operator that couples to the fluctuating force. Indeed, we have
\be
\Gamma_n = \sum_m \Gamma_{m \leftarrow n} 
= \int \frac{d\omega}{2\pi}
{\tilde{S}}(\omega) 
{\tilde{P}}(-\omega;E_n), 
\label{Gammatotalsimple}
\ee
with ${\tilde{P}(\omega;E_n) = \mbox{FT}[\langle \hat{\mathcal{A}}(t)\hat{\mathcal{A}}(0) \rangle]}$
that has the spectral decomposition  
\be
{\tilde{P}}(\omega;E_n) 
= \sum_m  \left| \left\langle  m \left| \hat{\mathcal{A}} \right| n \right\rangle  \right|^2      
2\pi  \delta(\omega-(E_m {-} E_n)). 
\label{simpleSystemSpectrum}
\ee
Note that ${\tilde{P}}$ obviously depends on the initial system state. 
The structure of Eq.~(\ref{Gammatotalsimple}) nicely indicates that each transition corresponds to extracting energy from the bath and putting it into the system or vice versa.

We now turn to the issue of dephasing. First we have to agree 
on a definition for the dephasing rate $\GammaSP$. 
The popular definition is based on a path integral approach (see App.\ref{sSC}),  
and it has two disadvantages: (i) it becomes ill defined 
outside of the semiclassical context; (ii) it involves 
an uncontrolled semiclassical (stationary phase) approximation 
which leads to results 
involving symmetrized rather than unsymmetrized spectral functions, 
which are thus not consistent with the FGR picture.
If one does not want to a adopt a context specific 
definition (e.g. relating to magnetoresistance)  
it is advantageous to define $\GammaSP$ as the decay 
rate of the purity~[\onlinecite{rgd,rgd2}] (see App.\ref{sPurity}), 
leading to a result that {\em does} agree with 
the heuristic FGR considerations which we clarify in the next paragraph, 
as well as with the more sophisticated diagrammatic approach.

One should be aware that the association between FGR transitions 
and decoherence is not strict for three reasons:  
{\bf \ (i)} Different preparations might have different rates 
of decoherence, and consequently there might be (say) 
two rather than one time constants. For example in NMR (see e.g.[\onlinecite{weiss}])
there is so called $T_1$ and $T_2$ time scales that describe 
the decay of vertical and horizontal components of the 
polarization vector;  
{\bf \ (ii)} A different, non FGR mechanism, might be involved.
For example in NMR the rate $1/T_2$ might have 
a contribution that comes from the so called ``pure dephasing" 
type processes which are related to energy levels fluctuating 
in time without inducing transitions between them. 
This contribution would be given by the fluctuation spectrum 
at zero frequency;         
{\bf \ (iii)}
Not {\em any} FGR transition implies decoherence, 
but only those that lead to entanglement and 
hence change both the purity of system and that of the environment.
This is further explained in App.\ref{sPurity} after Eq.(\ref{e18}).

With regard to (i) and (ii) we point out that 
for the physical system under study, namely, 
interacting electrons in a disordered metal  
in a thermal preparation, the rate $\GammaSP$ 
is assumed to be well defined in a statistical sense: 
there is no reason to assume multiple time scales,
or the existence of a rival mechanism of dephasing 
that comes from zero frequency fluctuations.  
With regard to (iii) we point out that in a more sophisticated treatment, 
using a diagrammatic approach, the elimination of the irrelevant
transitions is achieved by including ``vertex corrections”.
This leads to an effective {\em infrared} cut-off in the frequency
integral (\ref{Gammatotalsimple}), 
see for example Eq.(38) or (41) of [\onlinecite{munich1}].
However, our main concern here is to elucidate the role of Pauli
blocking, which will turn out to introduce an {\em ultraviolet} cutoff
into the frequency integral (\ref{Gammatotalsimple}). 
Thus, for the purpose of understanding the role of Pauli blocking, 
it is sufficient to ignore vertex corrections 
and to identify $\Gamma_\varphi$  as $\Gamma_n$ of Eq.(\ref{Gammatotalsimple}), 
appropriately averaged over the relevant energy window as
determined by the preparation or the temperature.

\section{The quantum noise approach applied to a many-body system}
\label{sec:qn-manybody}

As we have seen in the previous section,
it will be useful and instructive to write down the formula for the
dephasing rate of a particle interacting with an environment in terms
of an integral over a product of spectral functions that describe the
motion of the particle and the fluctuations of the environment.

In contrast to the preceding discussion, we now want to keep track of
the spatial degree of freedom explicitly, since it becomes relevant in
dephasing of particles moving in interferometers or a disordered
medium. More importantly, we also want to consider the general case of
a {\em many-body} system interacting with the environment. This will
enable us to automatically take into account the physics of Pauli
blocking which is crucial to correctly describe dephasing at low
temperatures.
The interaction between the particle(s)
and the environment will be written as
\beq{1}
{\hat V} \ \ = \ \  \int  \hat{\mathcal{U}}(x) \, \hat{\rho}(x) \, dx \,.
\eeq
Here the number density $\hat{\rho}(x)$ is either $\delta(x-\hat{x})$
for a single particle, or its many body version in general;  
while $\hat{\mathcal{U}}$ is the fluctuating potential, i.e. an operator 
which is associated with the environment. 
In the Heisenberg (interaction) picture a time index 
is added so we have $\hat{\mathcal{U}}(x,t)$ and $\hat{\rho}(x,t)$.

Following Ref. [\onlinecite{dph,dph2}], 
we define the spectral functions  $\tilde{S}$ and $\tilde{P}$ 
that characterize the fluctuations of the environment 
and the power spectrum of the motion, respectively:
\beq{2}
\tilde{S}(\bmq,\omega)
 &=& 
\iint
\Big[ \big\langle \hat{\mathcal{U}}(x,t)
\hat{\mathcal{U}}(0,0) \big\rangle \Big]
\, \eexp{i\omega t-iqx} \, dtdx,
\\ \label{e3}
\tilde{P}(\bmq,\omega)
 &=&  
L^d \iint
\Big[ \big\langle \hat{\rho}(x,t)
\hat{\rho}(0,0) \big\rangle \Big]
\, \eexp{i\omega t-iqx} \, dtdx.
\eeq
We assume stationary fluctuations for which 
the correlation functions depend only on the time 
and position differences. Unless otherwise specified 
the expectation value assumes a canonical (thermal) preparation. 
Note also that in Eq.(\ref{e3}) the total volume normalization 
with~$L^d$ is required in order to get expressions 
where the infinite volume limit is transparent. 
The spectral function $\tilde{S}(\bmq,\omega)$ is experimentally well defined:
it is essentially the {\em dynamic structure factor} 
(note remarks regarding notations in the last paragraph of this section). 
It is measurable in principle via scattering experiments, 
or via the dielectric function, to which it is related 
by the fluctuation-dissipation theorem: see App.\ref{sFDT}  
and in particular Eq.(\ref{e3008}) which describes 
the equilibrium fluctuations of the electrostatic potential within a dirty metal.  
Depending on the context, the physical identification 
of the spectral function $\tilde{P}(\bmq,\omega)$ 
might be more subtle, as discussed in subsequent sections.

We consider a situation where the many-body system and the environment are coupled weakly 
starting at $t=0$, and calculate the rate $\GammaSP$ for transitions induced 
by the coupling. As already explained in the previous section we identify 
this as the {\em dephasing rate}. As shown below, FGR leads 
to the following general result [\onlinecite{dph,dph2,imry,rgd,rgd2}],  
\beq{11} \label{GammaPhi}
\GammaSP \ \ = \ \  \int d\bmq \int
\frac{d\omega}{2\pi} \, \tilde{S}(\bmq,\omega) \,
\tilde{P}(-\bmq,-\omega), 
\eeq
which we call the ``SP formula" or the ``SP theory".  We stress that 
it is the {\em unsymmetrized} (quantum) versions of $\tilde{S}(q,\omega)$ and
$\tilde{P}(q,\omega)$ that enter this formula (see [\onlinecite{dph,dph2,asym2}]).  
The purpose of this paper is to discuss the implications of this statement in the many body context.

Equation (\ref{GammaPhi}) can be derived in the standard way. Consider the probability to have 
a transition induced by the system-environment coupling.  To lowest
order in the interaction (i.e. at short times), it reads
\beq{20}
p_{\varphi}(t)
\ \ &=& \ \
\int_0^t \int_0^t
\langle {\hat V}(t_2) {\hat V}(t_1) \rangle
\, dt_2dt_1
\\ \nonumber
\ \ &=& \ \
\iint dt_1dt_2 \iint dx_1 dx_2 
\\ \nonumber
&& \Big\langle  \hat \rho(x_2,t_2)
\hat{\mathcal{U}}(x_2,t_2) \, \hat{\rho}(x_1,t_1) \hat{\mathcal{U}}(x_1,t_1) \Big\rangle
\\
& = & 
\iint d\mathbf{q}\frac{d\omega}{2\pi} \tilde{S}(q,\omega)
\iint dt_1dt_2 
\iint dx_1 dx_2  \nonumber
\\ 
& &
\langle \hat \rho(x_2,t_2) \hat \rho(x_1,t_1) \rangle
\, \eexp{i\mathbf{q}\cdot (x_2-x_1)-i\omega(t_2-t_1)}.
\eeq
Employing the standard Golden Rule approach (i.e. going to the variables $(t_1+t_2)/2$ and ${\tau=t_2-t_1}$, 
and taking the appropriate limit of a total time span much larger than the correlation time), we obtain 
${p_{\varphi}\approx\GammaSP t}$, with $\GammaSP$ given
above in Eq.~(\ref{GammaPhi}).

All the standard caveats of this {\em linear-response} treatment apply. 
In particular, at very short times, $p_{\varphi}(t)$ will depend quadratically on time;  
during intermediate times there is a Wigner decay that agrees 
with FGR; and eventually (latest at the Heisenberg time)  
the perturbative picture breaks down.  
The condition of weak coupling which we impose, is tantamount to demanding 
that there is a large time-window between these two limits, during which the 
FGR-Wigner approximation is valid.

It is already apparent from the discussion in the previous section
that certain conditions have to be met in order to be able to identify
$\GammaSP$ of Eq.(\ref{e11}) as a dephasing rate for some meaningful, 
experimentally relevant observable. The situation we have in mind is that 
of a particle following different trajectories in an
interferometer or traveling through a disordered medium.
In a diagrammatic language, the loss of interference between 
these trajectories is given by Eq.(\ref{e11}) provided all 
induced transitions have comparable rates, and provided one is allowed 
to neglect so-called 'vertex corrections' [\onlinecite{munich1,munich2}]. 
While these vertex corrections are indeed important in describing, 
e.g., dephasing in weak localization, they are not our prime concern here.

We end this section with some brief remarks on notation. For
comparison, we indicate the relation to the notation adopted in
Ref.[\onlinecite{imry}].  
There it is assumed that the potential~$\mathcal{U}$
is induced by a background density~$\rho$.
The relation between the Fourier components of the potential 
and the Fourier components of the background density 
can be written as ${\mathcal{U}_{\bmq,\omega} = V_{\bmq,\omega} \rho_{\bmq,\omega}}$.
Assuming Coulomb interaction we have ${V_{\bmq,\omega}=4\pi e^2/q^2}$.  
Consequently ${\tilde{S}(\bmq,\omega) \equiv |V_{\bmq,\omega}|^2 S_{s}(\bmq,\omega)}$, 
where $S_{s}(\bmq,\omega)$ is identified as the {\em dynamic structure factor}. 
For the power spectrum of a single particle excitation Ref.[\onlinecite{imry}] 
has used the analogous notation $\tilde{P}^{[1]}(\bmq,\omega) \equiv S_{p}(\bmq,\omega)$.
These notations are oriented for the study of dephasing    
of electrons in dirty metals where the electrons are both
the ``system" and the ``bath" at the same time. 
Using these notations the ``SP formula" becomes:
\beq{666} 
\GammaSP^{[1]} =  \int d\bmq \int
\frac{d\omega}{2\pi} \, 
|V_{\bmq,\omega}|^2 \,
S_s(\bmq,\omega) \,
S_p(-\bmq,-\omega). 
\eeq
The spectral function $S_{s}(\bmq,\omega)$ 
is experimentally well defined as explained after Eq.(\ref{e2}).    
In contrast,
the object $S_{p}(\bmq,\omega)$ is a theoretical 
construction, motivated by a semiclassical picture (see next section) 
but not directly measurable.
In any case both spectral functions represent the ability 
of the bath and the system to {\em exchange energy and momentum}. 
Hence the physical meaning of Eq.(\ref{e666}) is quite transparent:  
It is the sum over all $(\bmq,\omega)$ modes that allow 
exchange momentum $\bmq$ and energy $\omega$ between 
the particle and the environment. The relative minus sign 
between the $(\bmq,\omega)$ arguments of the two spectral 
functions reflects the fact that energy (or momentum) taken 
by one is given to the other. Thus Eq.(\ref{e666}) is simply 
the total rate for exchanging "anything" between 
the particle and environment. It can be written down 
almost by inspection.

\section{The power spectrum $\tilde{P}(q,\omega)$}
\label{sPqomega}

The purpose of the present section is to calculate the spectral functions $\tilde{P}$
for the motion of non-interacting fermions in an arbitrary system.
We distinguish between single-particle and many-body 
spectral functions $\Pcl$ and $P^{[N]}$, 
describing the dynamics of some observable $\hat{A}$ 
(e.g. the density) in the context of single particle quantum mechanics  
or many-body Fermi~sea physics respectively.  
In particular, we show that the many-body spectral function $\tilde{P}^{[N]}$
can be written as the single-particle spectral function $\Pcl$ times the
number of fermions within the thermally smeared region around the Fermi surface.

\subsection{The single-particle spectrum $\Pcl$} 

We start with a very general discussion 
of spectral functions in non-interacting electronic systems. 
Given any single-particle observable~$\hat{A}$,  
we define the single particle power spectrum of~$\hat{A}$ 
in the absence of a Fermi sea as
\beq{0}
\Pcl(\omega;E)
&=& 
\mbox{FT} \Big[\bra \hat{A}(t) \hat{A}(0) \ket\Big]
\\ \nonumber
&=& 
\sum_{nm} p_n |A_{mn}|^2 \, 2\pi\delta(\omega - (E_m - E_n)),
\eeq
Here $p_n$ are the microcanonical weights which are peaked around $E_n \sim E$, 
namely ${p_n=\deltaF \times \delta(E_n-E)}$, where $\deltaF$ is the mean level spacing. 
It is convenient to take the bottom of the conduction 
band as the``potential floor" $E{=}0$.
Then if follows from the above spectral decomposition that
${\tilde{P}(\omega;E)=0}$ for ${\omega<-E}$, 
implying that the potential floor provides a lower cutoff for the emission tail. 
If~$E$ is well above the potential floor, than the
resulting spectrum $\Pcl$ is essentially classical, 
i.e. symmetric in~$\omega$ provided ${|\omega|\ll E}$.

It is implied by the definition Eq.(\ref{e3}) that 
the single-particle spectral function $\Pcl(q,\omega)$ 
is associated with the single-particle density 
operator ${\hat{A}=\delta(x-\hat{x})}$ 
or equivalently one may say that the fluctuations 
of the~$q$ Fourier component are associated 
with the special choice ${\hat{A}=\eexp{iqx}}$.  
As an important example (to be employed later on), 
we consider the motion of a single particle in a disordered potential. 
This motion is diffusive and accordingly
\beq{24}
\Pcl(q,\omega;E) \ \ = \ \ \frac{2Dq^2}{\omega^2+(Dq^2)^2},
\eeq
where $D$ is the diffusion coefficient.
Again we emphasize that it is implicitly assumed 
that the energy~$E$ of the particle 
is well above the potential floor,  
and we regard $D$ as a constant within the energy 
window of interest.

\subsection{The many-body spectrum $P^{[N]}$} 

\begin{figure}
\includegraphics[width=0.9\columnwidth]{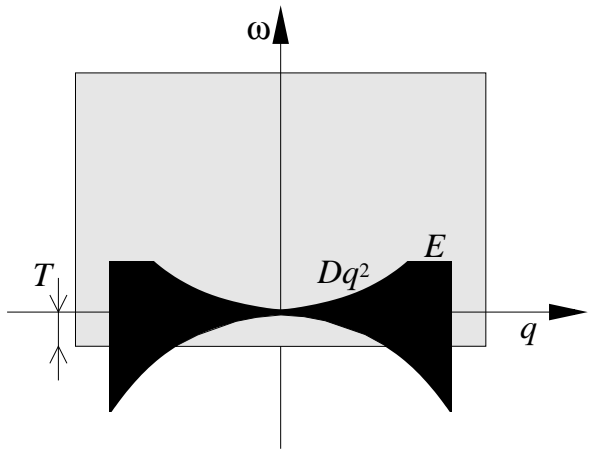}

\caption{\label{fqw} 
The $(q,\omega)$ plane.
The power spectra in a metallic system  
are distributed pre-dominantly 
within the shaded rectangular area that indicates an 
implicit momentum cutoff (inverse of the mean free path), 
and an implicit high frequency absorption cutoff 
(related to the rate of collisions). 
For the SP-formula it is essential to realize that 
the spectral function $\tilde{S}^{\tbox{eq}}(q,\omega)$
of Eq.(\ref{e3008}) has a lower emission cutoff 
which is determined by the temperature~$T$.
The power spectrum $\tilde{P}(-q,-\omega)$  
of either Eq.(\ref{e19}) or Eq.(\ref{e9}),  
which is associated with the diffusive motion 
of a particle, is concentrated pre-dominantly within 
the dark region $|\omega|\lesssim Dq^2$. 
For the SP-formula it is essential to realize 
that the energy~$E$ of the particle implies 
a frequency cutoff, which is analogous to~$T$. 
Close to equilibrium one should take ${E \sim T}$.}
\end{figure}

We now turn to the many-body spectral density. 
If we treat the many body system as a whole
then we have to employ second quantization to 
write ${\hat{\mathbf{A}} = \sum_{mn}A_{mn} a_m^\dagger a_n}$. 
Excluding the diagonal $n{=}m$ terms 
which are irrelevant to FGR transitions   
we get for a non-interacting system in a thermal state
\beq{19}
&\tilde{P}^{[N]}(\omega)&
\ \ = \ \ 
\mbox{FT} \bra \hat{\mathbf{A}}(t) \hat{\mathbf{A}}(0) \ket \label{manybodyPsp}
\\ \nonumber
&=&
\sum_{nm} |A_{mn}|^2 \bra a_n^\dagger a_m a_m^\dagger a_n\ket \, 2\pi\delta(\omega {-}(E_m{-}E_n))
\\ \nonumber
&=&
\sum_{nm} (1{-}f(E_m))f(E_n) |A_{mn}|^2 \, 2\pi\delta(\omega {-} (E_m {-} E_n))
\\ \nonumber
&=&
\int\frac{dE}{\deltaF} (1{-}f(E{+}\omega))f(E) \
\Pcl(\omega;E)
\\ \nonumber
&=&
\frac{\omega/\deltaF}{1-\eexp{-\omega/T}} \
\Pcl(\omega;E_{\tbox{F}}), 
\eeq
where the last expression is obtained if $\Pcl(\omega;E)$ 
is energy independent in the energy range of interest around the Fermi energy. 
Note that, at zero temperature, the first factor becomes $\deltaF^{-1} \omega \theta(\omega)$, cutting off all contributions at negative frequencies. Physically, this represents the fact that a zero-temperature fermion system can only absorb energy. If it couples to a zero-temperature environment, there will not be any transitions at all, in contrast to what would be deduced from the single-particle spectrum $\Pcl$ alone. The overlap between the power 
spectra $\tilde{S}$ and $\tilde{P}^{(N)}$ is illustrated in Fig.\ref{fqw}.

It is important to discuss the significance of the many body result.
To the extent that interactions can be neglected this 
result is {\em exact} and does not involve any uncontrolled 
semiclassical approximation. One realizes that  
\beq{170}
\tilde{P}^{[N]}(\omega) \ \ \approx \ \ \frac{1}{2} N_T \times \Pcl(\omega), 
\ \ \ \ \ \ \ \mbox{for $\omega {\ll}T$}, 
\eeq
where $N_T=2T/\deltaF$ is the effective number of particles 
in the thermally smeared band around the Fermi surface.
This factor is extensive, i.e. it grows linearly with the system's volume.
Only these particles can be excited by a small amount 
of energy $\omega\sim0$ being absorbed from the environment, 
or, vice versa, they can release some energy into the environment. 
It is crucial to keep the physics of Pauli blocking correctly in this description. 
In fact, had we neglected the Pauli blocking, 
instead of Eq.(\ref{e170}) we would
have obtained  ${\tilde{P}^{[N]}(\omega)= N \Pcl(\omega)}$,  
where $N$ would have indicated the total number of particles in the system.

\subsection{The single excitation spectrum $P^{[1]}$} 

In the present subsection we would like to 
make contact with other descriptions in the literature,
where the focus is on the dephasing of 
a single-particle excitation in the presence of a Fermi sea.
Unlike the many-body calculation above, 
such point of view requires to introduce Pauli blocking 
factors "by hand". One finds (see Eq.(\ref{e9}) below)  
that the power spectrum of a thermalized one-particle excitation is   
\beq{171}
\tilde{P}^{[1]}(\omega) \ \ \approx \ \ \frac{1}{2} \times \Pcl(\omega),
\ \ \ \ \ \ \ \mbox{for $\omega {\ll}T$},
\eeq
which implies that Eq.(\ref{e170}) can be re-written 
as ${\tilde{P}^{[N]}(\omega) \approx N_T \tilde{P}^{[1]}(\omega)}$,  
which holds in the vicinity of the Fermi energy.
One realizes that the factor $1/2$ reflects that Pauli blocking 
of the downward transitions and persists at high temperatures.  
The relation between the spectrum of the single-particle motion and the many-body 
density fluctuations is a central result of the present section. 
It will be employed in the next section to connect the one-body and the many-body dephasing rates.

In the spirit of the prevailing literature we consider separately 
electrons and holes of arbitrary energy, incorporating Pauli blocking 
factors ``by hand'' into the definition of the particle's power spectrum. 
For an electron above the Fermi sea it has been claimed in
Ref.[\onlinecite{dph,dph2}] that the Fermi energy~$E_F$ is like an
effective potential floor. This is implied by the Fermi statistics,
taking into account that a one-body-operator can change the state of
only one electron. It also results from a more detailed analysis of
the consequences of Pauli blocking on dephasing, which has been
carried out both in the context of weak localization
[\onlinecite{munich1,munich2}] and ballistic interferometers
[\onlinecite{2005_Marquardt_MZQB_EPL,2006_04_MZQB_Long}], and which
has been illustrated in exactly solvable models as well
[\onlinecite{2004_Marquardt_DFS_PRL,2004_Marquardt_DFS_LongVersion}].
Consequently we define the power spectrum of an electron excitation
with the appropriate Pauli blocking factors incorporated:
\beq{-1}
\tilde{P}^{[e]}(\omega;E)
&=& 
\sum_{nm} (1{-}f(E_m))p_n |A_{mn}|^2 \, 2\pi\delta(\omega {-} (E_m {-} E_n))
\\
&=& 
(1-f(E{+}\omega)) \ \Pcl(\omega;E).
\label{e191}
\eeq
An analogous expression is introduced for holes:
\beq{-1}
\tilde{P}^{[h]}(\omega;E)
&=&
\sum_{nm} f(E_m) p_n |A_{mn}|^2 \, 2\pi\delta(\omega {-} (E_n {-} E_m))
\\
&=& 
f(E{-}\omega) \ \Pcl(\omega;E).
\label{e192}
\eeq
We can thermally average over $E$ using the prescription 
\beq{0}
\overline{G(E)} \ \ \equiv \ \ \int_{-\infty}^{\infty}  G(E) \ [-f'(E)] dE 
\eeq
where $f(E)$ is the Fermi occupation function which 
is determined by the Fermi energy $E_{\tbox{F}}$ 
and the temperature~$T$. Then we get 
\beq{9}
\tilde{P}^{[1]}(\omega) 
\ \ &\equiv& \ \
\overline{\tilde{P}^{[e]}(\omega;E)}
\ \ = \ \
\overline{\tilde{P}^{[h]}(\omega;E)}
\\ \nonumber
\ \ &=& \ \
\frac{d}{d\omega}\left[\frac{\omega}{1-\eexp{-\omega/T}}\right] \times \Pcl(\omega;E_{\tbox{F}}).
\eeq
Note that, at zero temperature, the first factor is just a step-function $\theta(\omega)$, cutting off the contributions from negative frequencies.

\subsection{Density fluctuations} 

By specializing the above general discussion 
to the case that $\hat{A}$ represents the density operator, 
and using the reasoning of the present section, we can define 
the $\tilde{P}^{[e]}(q,\omega;E)$ of an electron in the Fermi sea;
the $\tilde{P}^{[h]}(q,\omega;E)$ of a hole in the Fermi sea;
the $\tilde{P}^{[1]}(q,\omega)$ of a single particle excitation 
at equilibrium; and the $\tilde{P}^{[N]}(q,\omega)$ 
of the whole many body electronic system.

The results Eq.(\ref{e9}), Eq.(\ref{e191}), Eq.(\ref{e192}) and Eq.(\ref{e19}), 
as well as the approximations Eq.(\ref{e171}) and Eq.(\ref{e170}) 
for the power spectrum of the system are all of the form 
\beq{455}
\tilde{P}(q,\omega) \ \ = \ \ f_p(\omega) \, \Pcl(\bmq,\omega). 
\eeq
where $f_p(\omega)$ reflects the way in which 
the Fermi occupation statistics manifests itself.

\section{Dephasing of a many-fermion system for weak coupling to the environment}
\label{sMetal}

The purpose of this section is to work out the many-body dephasing rate $\GammaSP^{[N]}$ according to SP-theory, and to compare it with the single-particle dephasing rate $\GammaSP^{[1]}$ that is obtained after incorporating Pauli blocking factors into the spectra, as described in the previous section. We will see they are equal, up to a factor describing the number of thermally excited particles that effectively can participate in the processes leading to dephasing.

Before we derive the dephasing rate, we set up a few simplifications in the notation.
The integrand of the ``SP formula" Eq.(\ref{e11}) includes 
a product of two spectral functions. 
We assume that the ``bath" is in thermal equilibrium 
and therefore the detailed balance condition 
that should be satisfies is:
\beq{0}
\frac{\tilde{S}(q,\omega)}{\tilde{S}(q,-\omega)} \ \ = \ \ \eexp{\omega / T}.
\eeq
It follows that the spectral function can be written as 
\beq{454}
\tilde{S}(q,\omega) \ \ = \ \ \left[\frac{2\omega}{1-\eexp{-\omega/T}}\right] \, \eta(q,\omega), 
\eeq
where $\eta(q,\omega)$ is a symmetric function. 
It represents a generalized friction
coefficient (in analogy to the standard notation in the Caldeira-Leggett model).
Note that at high temperatures we have ${\tilde{S}(q,\omega)=2\eta T}$.

Without any approximation involved, because 
the $d\omega$ integration extends from $-\infty$ to $\infty$, 
the integrand of the ``SP formula" 
can be symmetrized using the replacement 
\beq{0}
F(\omega)  \ \ \mapsto \ \ 
\{F(\omega)\}_{\tbox{sym}} \equiv 
\frac{1}{2}[F(\omega)+F(-\omega)].
\eeq
It is now natural to combine the prefactors 
in Eq.(\ref{e454}) and Eq.(\ref{e455}) 
into a ``frequency-dependent temperature" 
for $\omega$ transitions:
\beq{0}
T(\omega)  \ \ \equiv \ \ 
\left\{ 
\left[\frac{2\omega}{1-\eexp{-\omega/T}}\right]    
f_p(-\omega)
\right\}_{\tbox{sym}}. 
\eeq
and to define an associated symmetrized spectral function 
for the effective thermal fluctuation of the environment:  
\beq{0}
\tilde{S}^{(0)}(\bmq,\omega) \ \ \equiv \ \ 2\eta(q,\omega) T(\omega).
\eeq
These definitions allow the ``SP formula" to be written 
in a symmetrized form that  involves a product of functions 
that are symmetric in~$\omega$. Namely, 
\beq{240} 
\GammaSP= \frac{1}{2}\int d\bmq \int
\frac{d\omega}{2\pi} \, \tilde{S}^{(0)}(\bmq,\omega)  \, \Pcl(\bmq,\omega). 
\eeq

In the following subsections we shall discuss the functional form 
of $T(\omega)$ which is crucial for the calculation 
of low temperature dephasing.  But first let us 
illuminate the outcome of Eq.(\ref{e240}) in 
what can be termed the semiclassical Nyquist limit.
Namely, considering high temperatures, 
for which not only $\Pcl$ but also $\tilde{S}^{(0)}$ 
are classical-alike, one realizes that Eq.(\ref{e240}) 
still contains a non-classical~$1/2$ 
due to the Pauli blocking of the downward transitions.
So strictly speaking Eq.(\ref{e240}) does not possess 
a classical limit.   
If  $\eta$ is independent of~$\omega$ and the motion 
of the particle involves only small 
frequencies ${|\omega| \ll T}$, then~$T^{[1]}(\omega)$ 
in the integrand of Eq.(\ref{e240})
can be replaced by the temperature~$T$, 
and we get the simple result
\beq{0} 
\GammaSP^{[1]} \ \ = \ \ \alpha T,
\eeq
where the dimensionless $\alpha$ is the $dq$ integral over $\eta$. 
But if we consider (say) a diffusive electron,  
then at low temperatures its power spectrum 
is broader than $T$ if ${q>(T/D)^{1/2}}$, 
as illustrated in Fig.\ref{fqw}. 
Then the weight which is provided by~$T(\omega)$ 
is like an effective cutoff, leading to non-linear   
dependence on the temperature. See e.g. [\onlinecite{dph,dph2}] 
for a gallery of various results.

\subsection{The dephasing rate of quasi particles at equilibrium}

The calculation of the one body 
dephasing rate $\GammaSP^{[1]}$ involves
the spectral function $\tilde{P}^{[1]}(\bmq,\omega)$ 
of Eq.(\ref{e9}) and hence  
\beq{0} 
T(\omega)^{[1]} \equiv 
\left\{
\left[\frac{2\omega}{1-\eexp{-\omega/T}}\right]
\frac{d}{d\omega}
\left[\frac{\omega}{1-\eexp{\omega/T}}\right]
\right\}_{sym}.
\eeq
The calculation of the $N$ body 
dephasing rate $\GammaSP^{[N]}$ involves 
the spectral function $\tilde{P}^{[N]}(\bmq,\omega)$ 
of Eq.(\ref{e19}) and hence 
\beq{0} 
T(\omega)^{[N]} \equiv 
\left\{
\left[\frac{2\omega}{1-\eexp{-\omega/T}}\right]
\left[\frac{-\omega/\deltaF}{1-\eexp{\omega/T}}\right]
\right\}_{sym}.
\eeq
Doing the algebra we get
\beq{0} 
T(\omega)^{[1]} \ \ = \ \  
\left[\frac{(\omega/2T)}{\sinh(\omega/2T)}\right]^2 \, T,
\eeq
and $T(\omega)^{[N]} = (2T/\deltaF)T(\omega)^{[1]}$, 
%
%
leading to 
\beq{29}
\GammaSP^{[N]} \ \ = \ \ \left(2\frac{T}{\deltaF}\right) \GammaSP^{[1]}. 
\label{CentralResult}
\eeq
We have thus come to the conclusion that the many-body dephasing rate
$\GammaSP^{[N]}$ is equal to the single-excitation dephasing rate
$\GammaSP^{[1]}$ which properly incorporates Pauli blocking,
multiplied by a factor $N_T=2T/\deltaF$ that counts the number of thermally
excited particles.
At this point it is important to emphasize that if the interfering entity (system) 
consists of $N_s$ constituent particles, all interacting in the same way with the environment, 
one expects the dephasing rate to be $N_s$ times the dephasing rate 
for a single constituent particle under the same conditions. 
This is why larger systems are expected to decohere faster [\onlinecite{Ze1,Ze2}].  
In our case the effective number of participating particles 
in the system is $N_T$ irrespective of the actual total number. 
Thus, without putting in "by hand" any Pauli blocking factors, 
we have re-derived the correct
result that is obtained from a more sophisticated diagrammatic or
path-integral analysis [\onlinecite{munich1,munich2}].  It is
therefore possible to regard $\GammaSP^{[1]}$ as the dephasing rate
per effective particle. Equation (\ref{CentralResult}) represents 
a central result of the present paper, whose consequences and
modifications will be discussed in the following.

\subsection{The dephasing rate of a non-equilibrium excitation}

It is interesting as well to consider the dephasing rate for a
non-equilibrium one-particle excitation at some energy $E$. Then we
cannot simply start from the many-body spectrum $\tilde{P}^{[N]}$
which is calculated for the thermal equilibrium state of the many-body
problem. To make contact with previous approaches in the literature
[\onlinecite{munich1}], we briefly formulate the
nonequilibrium dephasing rate in terms of the present notations.
In Ref.[\onlinecite{munich1}] it has been argued that the dephasing rate
should be calculated as
\beq{0}
\GammaSP^{\tbox{noneq}}(E)
= \frac{1}{2}\left(\GammaSP^{[e]}(E) + \GammaSP^{[h]}(E)\right).
\eeq
If we make a thermal average over $E$ both terms are equal,
but if we consider non-equilibrium excitations a more
careful treatment is required.  
Using the expressions for ${\tilde{P}^{[e]}(q,\omega;E)}$
and ${\tilde{P}^{[h]}(q,\omega;E)}$, 
the corresponding function in Eq.(\ref{e455}) is
\beq{0}
f_p(\omega) = [1-f(E{+}\omega) + f(E{-}\omega)]/2,  
\eeq
hence
\beq{0}
T(\omega) = 
\left\{
\left[\frac{\omega}{1{-}\eexp{-\omega/T}}\right] \Big(1{-}f(E{-}\omega){+}f(E{+}\omega)\Big)
\right\}_{sym}.
\eeq
Doing the algebra we get
\beq{-1} 
T(\omega) =
\left[ \left(\frac{1}{\eexp{\omega/T}{-}1}\right) +
\frac{1}{2}\Big(1{-}f(E{-}\omega)+f(E{+}\omega)\Big)\right]
\omega 
\\ \nonumber
= 
\left[\coth\left(\frac{\omega}{2T}\right)
+\frac{1}{2}\tanh\left(\frac{E{-}\omega}{2T}\right)
-\frac{1}{2}\tanh\left(\frac{E{+}\omega}{2T}\right) \right]
\omega,
\eeq
which agrees with Ref.[\onlinecite{munich1}].

\section{Dephasing of electrons with screening: The diagrammatic perspective }
\label{sDiagrm}

Up to now, we have treated the weak-coupling case, in which neither
the particle dynamics (i.e. the density fluctuations) nor the potential
fluctuations are influenced appreciably by the presence of interactions.
It is for this reason that we have been able to derive the total particle
decay rate in such a simple manner, expressing it through the product
of correlators $\tilde{P}$ and $\tilde{S}$, associated with the
particle dynamics and the potential fluctuations, respectively.

More care has to be exercised when one tries to extend the framework
to cases in which the potential fluctuations themselves change strongly
(even qualitatively) after switching on the interaction. The most
important example concerns the effect of the long-range Coulomb interaction
in a metal. The long-range repulsion suppresses efficiently long-wavelength,
low-frequency density fluctuations and the associated potential fluctuations,
thus giving rise to screening. An important counter-intuitive (yet
well-known) consequence is that the potential spectrum at small $\omega,q$
is independent of the electron charge (see Eq.(\ref{e3008})).  

It will turn out that, for such a situation, the single-particle decay rate is not correctly
reproduced with a naive ansatz, in which both $\tilde{S}$ and $\tilde{P}$
are obtained for the full interacting system (neither, of course,
can we neglect interactions completely in both $\tilde{S}$ and $\tilde{P}$).
To shed light on this issue, we now rephrase the results of the ``SP theory''
developed here in terms of diagrams.

\begin{figure}
\includegraphics[width=\columnwidth]{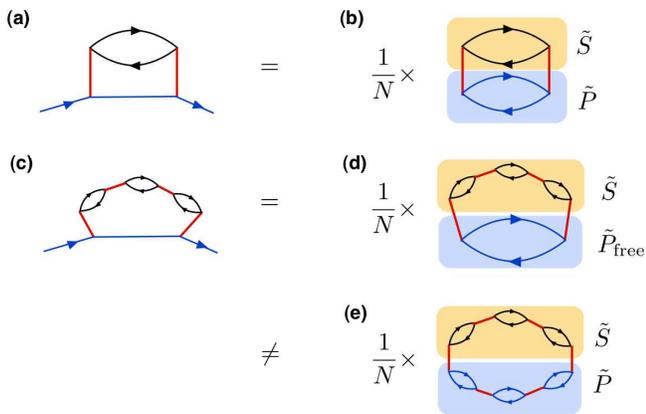}

\caption{\label{fDiagrams}
The relation between ``SP theory'' and standard diagrammatics: 
(a) Diagrammatic calculation of the single-particle
decay rate to leading order in the density-density interaction. 
(b) When applying the ``SP theory'' to a many-body system, one
effectively calculates an extensive diagram, 
whose value scales with the number of affected particles 
(here denoted as ``$N$''). 
(c) For a metal with screening, diagrams of this type have to be summed
to yield the single-particle decay rate in RPA. 
(d) After translation into the diagrams appearing 
in the ``SP theory'', we see that the bare particle-hole bubble 
has to be used in describing the system's motion, in contrast to (e). }
\end{figure}

\subsection{Diagrams for the weak-coupling limit}

We first return to the weak-coupling case. In that limit, the single-particle
decay rate is obtained in a diagram of the type shown in Fig.\ref{fDiagrams}(a).
It represents the interaction of the given particle with the density-fluctuations
described by the bubble. Turning to SP-theory, we see the
following: Its straightforward application to a many-body system produces
a diagram of the type shown in Fig.\ref{fDiagrams}(b). This
diagram has no external lines, and is therefore extensive, i.e. its
value grows with volume. More precisely, as elaborated in our previous
discussion, it yields the decay rate multiplied by the number of particles
that can be scattered. Once this feature is taken into account, one
can deduce the single-particle decay rate.

\subsection{Diagrams beyond weak-coupling}

Let us now have a look at the strong-coupling case, where screening
alters drastically the fluctuations of the density and the potential.
Within the random phase approximation (RPA), the single-particle decay
rate can be calculated using diagrams of the type shown in Fig.\ref{fDiagrams}(c),
with an arbitrary number of polarization bubbles inserted to account
for screening. In this way, the correct modified fluctuation spectrum
$\tilde{S}$ enters the decay rate. When translating this into an
appropriate extensive diagram (Fig.\ref{fDiagrams}(d)), the
open-ended single-particle line turns into one particle-hole bubble.
The latter corresponds to the density-density correlator \emph{evaluated
in the absence of interactions}. In contrast, in a literal (naive)
application of SP-theory to the many-body problem with screening
one might be tempted to employ the screened density-density propagator
(Fig.\ref{fDiagrams}(e)), which gives incorrect results.

The source of the difficulty is quite obvious when 
comparing Fig.~\ref{fDiagrams}(e) to Fig.~\ref{fDiagrams}(c): 
The origin of the bubble is the single-particle
line, which appears since we are interested in the coherent propagation
of the particle. We are not interested in the propagation of a density
perturbation that enters diagram (e).

The issue discussed here is independent of whether we are in the ballistic
or the diffusive regime, or of what are the details of the model.
In any case, the correct application of the SP-theory requires  
to employ the full potential fluctuations (including screening)
for $\tilde{S}$, while looking at the bare $\tilde{P}_{{\tbox{free}}}^{[N]}$,
calculated for the non-interacting system.

\section{Self consistent point of view for the dephasing of interacting electrons}
\label{sInteract}

\begin{figure}
\includegraphics[width=\columnwidth]{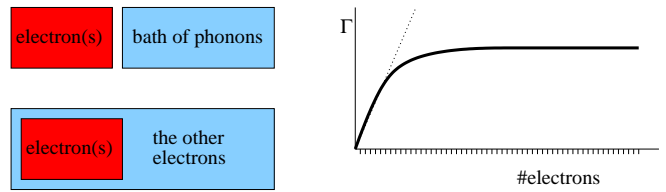}

\caption{\label{fep} 
We first consider non-interacting electrons 
in a box\cite{rmA} of volume~$L^d$ coupled 
to a bath of phonons at temperature~$T$.  
The solid curve on the right illustrates 
how the decay rate of the {\em purity}  
depends on the number of electrons in the system.   
If the electrons were not identical (dotted line) 
one would get ${\Gamma=N\Gamma^{[1]}}$, 
but due to the Fermi statistics one obtains 
a saturation at ${\Gamma=N_T\Gamma^{[1]}}$. 
In the case of $N$ interacting electrons 
in a dirty metal, with no bath of phonons involved,  
one can define a bunch of $N_s$ test particles 
as the {\em system}\cite{rmB},  
while all the other $N{-}N_s$  
electrons constitute the fluctuating environment. 
We have in mind ${N_s \ll N}$, while for ${(N{-}N_s) \ll N}$, 
the role of system and environment is flipped.
}
\end{figure}

We now want to clarify within 
the SP~approach how screening should be handled,  
and demonstrate the consistency with 
the diagrammatic perspective of Section~\ref{sInteract}.
In a dirty metal the motion of electrons is diffusive, 
and there is a screening effect due to the 
long range Coulomb interaction. The fluctuations 
of the electrostatic potential $U$ reflect 
the many-body fluctuations in the density 
of the electrons via the Coulomb law
\beq{111}
\tilde{S}(q,\omega) \ \ = \ \ \frac{1}{L^d}\left(\frac{4\pi e^2}{q^2}\right)^2 \ \tilde{P}(q,\omega),
\eeq
with $\tilde{S}$ and $\tilde{P}$ defined by Eqs.~(\ref{e2}) and~(\ref{e3}) respectively. 
The equilibrium fluctuations of the induced electrostatic fluctuations $\tilde{S}^{\tbox{eq}}(q,\omega)$
are trivially related to the dynamic form factor $\tilde{P}^{\tbox{eq}}(q,\omega)$ 
by the above relation, and can be determined self consistently using the 
fluctuation-dissipation relation (see App.\ref{sFDT}):  
\beq{0}
\tilde{S}^{\tbox{eq}}(q,\omega) \ \ = \ \ 
\frac{1}{\nu Dq^2} 
\left[\frac{2\omega}{1-\eexp{-\omega/T}}\right].
\eeq
If we want to calculate the single-particle dephasing rate  $\GammaSP^{[1]}$, 
then one option is to obtain $\tilde{P}^{[1]}(q,\omega)$ from a blend 
of semiclassical and many-body considerations as in Section~\ref{sPqomega}.
But we are trying in this paper to explore an alternative route, 
where the single particle dephasing rate 
is defined as the many-body dephasing rate $\GammaSP^{[N]}$  
per effective particle as in Eq.(\ref{CentralResult}).
The main difficulty that immediately arises,  
once interacting electrons are concerned, 
is how to define the ``system" for which the calculation is done. 
The proper formulation of the system-bath paradigm 
becomes tricky once electrons are both 
the ``system" and the ``bath" (see illustration in Fig.\ref{fep}).  
In particular we have to determine what is $\tilde{P}^{[N]}(q,\omega)$
for the case that screening is important.
Here we encounter a conflict between two opposing points of view, 
which we discuss below.

There are two ways to determine $\tilde{P}^{[N]}(q,\omega)$. 
{\em On the one hand}, 
within the framework of the heuristic semiclassical approach of Section~\ref{sPqomega}, 
the spectral function $\tilde{P}^{[N]}(q,\omega)$  
should be given by Eq.(\ref{e19}) which treats 
the electrons as non-interacting. 
{\em On the other hand}, 
within the formal framework it describes the fluctuations 
of the many body density of the electrons, 
and therefore should equal $\tilde{P}^{\tbox{eq}}(q,\omega)$ 
as defined after Eq.(\ref{e111}).
But then one realizes that for an {\em interacting} system  
$\tilde{P}^{[N]}(q,\omega)$ is very different 
from $\tilde{P}^{\tbox{eq}}(q,\omega)$. 
So it seems that we have here an inconsistency.

In order to resolve the apparent inconsistency  
we have to clarify the physical meaning of 
the phrase {\em ``dephasing rate per particle"}. 
This notion is not problematic conceptually if the environment 
is a distinct entity (phonon bath). 
But in the case of a dirty metal this distinction is blurred: 
obviously we cannot regard the same particles as both the ``system" 
and ``environment".

It turns out that a reconciliation of the formal 
fluctuation-dissipation analysis with the heuristic 
approach is possible provided we define the ``system" 
as a bunch of non-interacting test particles\cite{rmC},    
whose density fluctuations are simply those of diffusing, 
non-interacting electrons. We can obtain the power-spectrum 
of the `minority' test particles, starting from Eq.(\ref{e98}), 
but replacing $U_{\tbox{total}}$ by the fluctuating potential~$\mathcal{U}$  
produced by the 'majority' particles. This effectively omits screening effects 
in the calculation of the test particle density fluctuations:
\beq{435}
\rho_{\bmq,\omega} \ \ = \ \ \frac{(\sigma/e^2) q^2}{i\omega-Dq^2} \ \mathcal{U}_{\bmq,\omega}.
\eeq
Note that the $\rho$ of the system (minority fraction of electrons) 
is assumed to be much smaller than the total electronic 
density $\rho_{\tbox{elct}}$ that appears in the FDT derivation, 
hence one can neglect the back-reaction effect. In other words, 
as far as the minority $\rho$ is concerned we do not take the 
screening into account, and treat them as non-interacting. 
From Eq.(\ref{e435}), together with Eqs.(\ref{e2}) and (\ref{e3}) 
and the Einstein relation $\sigma= e^2 D \nu$, we get immediately 
\beq{222}
\tilde{P}^{[N]}(\bmq,\omega)
\ \ &=& \ \
L^d \, 
\left|\frac{\nu D q^2}{i\omega-Dq^2} \right|^2  
\ \tilde{S}^{\tbox{eq}}(\bmq,\omega)
\\
\ \ &=& \ \
\frac{\omega/\deltaF}{1-\eexp{-\omega/T}} \
\Pcl(\omega;E_{\tbox{F}}), 
\eeq
where for the second line we recalled Eq.~(\ref{e24}) for $\Pcl$ 
and Eq.~(\ref{e3008}) for $\tilde{S}^{\tbox{eq}}$. 
This result agrees with Eq.(\ref{e19}) as if we could regard 
the electrons as diffusing but non-interacting. 
It is important to appreciate that we have obtained 
here a non-trivial profound relation that {\em bypasses the heuristic approach} 
which is required in order to adopt Eq.(\ref{e19}) 
for strongly interacting electrons: strictly speaking the derivation 
that leads to Eq.(\ref{e19}) is not applicable here. 
Still we get Eq.(\ref{e222}) which agrees with Eq.(\ref{e19}) 
by extending the common fluctuation-dissipation reasoning, 
without the need to introduce a blend of semiclassics with Pauli exclusion factors.

\section{Conclusions}

We have presented a straightforward approach to the
calculation of the dephasing rate within the framework 
of Fermi golden rule picture, and applied it to a many-fermion system. 
Starting from the quantum spectra of the environment ($\tilde{S}$) and the system ($\tilde{P}$), the
approach (termed here "SP-theory") yields the dephasing rate as 
an integral over the frequency transferred between system and environment
during interaction processes. In the present paper, we have gone beyond previous
attempts, and considered a full many-fermion system. We have argued
that this yields results which automatically incorporate the crucial physics of
Pauli blocking that serves to suppress decoherence at low temperatures. 
The many-body dephasing rate can be identified as the single-particle
dephasing rate times the effective number of thermally excited particles susceptible to scattering.
The use of non-symmetrized spectral functions provides 
in a natural way the proper cutoff scheme 
for the frequency integral of the "SP formula", 
without having to invoke the ad hoc
cutoff schemes that had been introduced in previous publications.

By defining the single particle dephasing as the many-body 
dephasing per particle one can bypass the need to use 
a blend of semiclassics with Pauli exclusion factors.
This point of view also provides a natural bridge  
to the diagrammatic approach. 
Indeed we have shown how the results of the SP-theory 
can be interpreted in terms of diagrams. 
This has allowed us to address another question, namely how SP-theory 
should be applied in a situation in which the system-environment coupling is no longer weak.
That is the situation relevant for electrons moving in a disordered metal, where screening
is crucial for the structure of the Nyquist noise. 
We have shown that SP-theory should incorporate 
in such a case the full environment spectrum, 
alongside the non-interacting density spectrum of the system, 
in agreement with the diagrammatic approach.


\ \\

\noindent
{\bf Acknowledgment:} 
We thank Ora Entin-Wohlman and Baruch Horovitz for useful discussions.
The research has been supported by a grant from the DIP, the
Deutsch-Israelische Projektkooperation (contract H-2-1), 
as well as by the DFG through the Emmy-Noether program (FM) and SFB TR 12. This research was
supported in part by the National Science Foundation under Grant
No. NSF PHY05-51164, and by the USA-Israel Binational Science Foundation (BSF).

\appendix

\section{The semiclassical picture of dephasing}
\label{sSC}

The notion of dephasing naturally arises in the 
analysis of transport where, loosely speaking, 
one is interested in calculating the probability 
of a particle to get from one point to a different point.
Consequently the most popular definition of the 
dephasing factor is based on a semiclassical picture.   
Using the Feynman-Vernon formalism, 
and adopting the notations as in Ref.[\onlinecite{dph,dph2}],  
the propagator of the reduced probability matrix 
in the presence of a thermal bath, is expressed
as a sum over pairs of classical trajectories:
\beq{0}
\sum_{ab} A_aA_b^{*} 
\exp\left(-\frac{S_N[x_a,x_b]}{\hbar^2}\right) \ 
\exp\left(i\frac{S[x_a]{-}S[x_b]}{\hbar}\right).
\eeq
The dephasing factor $P_{\varphi}(t)$ is defined 
as a number within $[0,1]$ that
characterizes the suppression of the off-diagonal elements: 
One observes that after time~$t$ the interference contribution,
that comes from the off diagonal terms in the double sum, 
is suppressed by a factor 
\beq{16}
P_{\varphi}(t)
= \eexp{-S_N[x^A,x^B]}
\equiv 
\Big| \ \Big\langle \ U[x_a] \chi \ \Big| \ U[x_b] \chi \ \Big\rangle \ \Big|,
\eeq
where $\chi$ is the preparation of the bath.  In order not to
complicate the notations, the canonical average over the $\chi$ states
is implicit.  The unitary operator $U[x]$ generates the evolution of
the bath given that the particle goes along the trajectory~$x(t)$. The
action $S_N[x_a,x_b]$ is a double time integral. Using manipulation as
in Refs.~[\onlinecite{dph,dph2}] one obtains the ``SP formula" with
the symmetrized version of $\tilde{S}(q,\omega)$, and the symmetric
classical version of $\tilde{P}(q,\omega)$, namely $\Pcl$ of Eq.(\ref{e24}).

The semiclassical expression is definitely wrong for short range scattering 
at low temperatures [\onlinecite{dph,dph2}], because it does not reflect 
that {\em closed channels cannot be excited}. 
This problem with the semiclassical (stationary phase) approximation 
is well known in the theory of inelastic scattering.

\section{The purity based definition of dephasing}
\label{sPurity}

The following appendix follows the presentation 
of Ref.~[\onlinecite{rgd,rgd2}]. The natural definition 
for the dephasing factor $P_{\varphi}(t)$  is related 
to the purity $\trc(\rho^2)$ of the reduced
probability matrix. Given that the state of the
system including the environment is $\Psi_{pn}$, where $p$ and $n$
label the basis states of the particle and the bath respectively,
the full probability matrix is $\Psi_{pn}\Psi_{p'n'}^*$, 
while the purity of the {\em reduce} probability matrix is
\beq{17}
P_{\varphi}(t)
&=& 
\sqrt{\trc(\rho_{\tbox{sys}}^2)}
= 
\sqrt{\trc(\rho_{\tbox{env}}^2)}
\nonumber
\\
&=&
\left[\sum_{p'p''n'n''} \Psi_{p'n'}\Psi_{p''n'}^*\Psi_{p''n''}\Psi_{p'n''}^*\right]^{1/2}.
\eeq
Assuming a factorized initial preparation as in the conventional
Feynman-Vernon formalism, we propose the rate of loss of purity 
as a measure for decoherence. A standard reservation applies:
initial transients during which the system gets ``dressed" by the environment
should be ignored as these reflect renormalizations
due to the interactions with the high frequency modes.
Other choices of initial state might involve
different transients, while the later slow approach
to equilibrium should be independent of these transients.
In any case the reasoning here is not much different
from the usual ideology of the Fermi golden rule,
which is used with similar restrictions to calculate
transition rates between levels.

Writing the initial preparation as ${\Psi^{(0)}_{pn}=\delta_{p,p_0}\delta_{n,n_0}}$,
and using leading order perturbation theory, we can relate $P_{\varphi}$ to the
probabilities ${P_t(p,n|p_0,n_0)=|\Psi_{pn}|^2}$ to have a transition
from the state $|p_0,n_0\rangle$ to the state $|p,n\rangle$ after time~$t$.
The derivation is detailed in Appendix~E of Ref.[\onlinecite{rgd,rgd2}]. 
One obtains the result
\beq{18}
P_{\varphi}(t) &=& P_t(p_0,n_0|p_0,n_0) 
\\ \nonumber 
&+& P_t(p{\ne}p_0,n_0|p_0,n_0) +  P_t(p_0,n{\ne}n_0|p_0,n_0).
\eeq
The notation $p\neq p_0$ or $n\neq n_0$ implies a
summation $\sum_{p\neq p_0}$ or $\sum_{n\neq n_0}$, respectively.
The actual calculation of $P_t(p,n|p_0,n_0)$ can be done 
using Fermi's golden rule (FGR) as discussed in the main text.

Thus we see that within the FGR framework, the purity 
is simply the probability that either the system or the bath 
do not make a transition. Accordingly $P_{\varphi}(t)$ 
is essentially the same as the survival probability $P(t)$
of the initial state (the first term in Eq.(\ref{e18})).
In typical circumstances the difference 
between $P_{\varphi}(t)$ and $P(t)$ has zero measure weight 
in the $dq d\omega$ integration and therefore $\GammaSP$ can be 
identified with the Wigner decay rate of system excitations.

\section{The fluctuation-dissipation relation}
\label{sFDT}

The dielectric constant of a metal is defined via the linear relation
between the total electrostatic potential $U_{\tbox{total}}$ 
and an external test charge density $\rho_{\tbox{ext}}$  
\beq{97}
U_{\tbox{total}} \ \ = \ \ \frac{1}{\varepsilon(q,\omega)} \left(\frac{4\pi e^2}{q^2}\right) \rho_{\tbox{ext}}.
\eeq
For simplicity we relate here and below to one component $q$ of the fields. 
The total electrostatic potential is the sum 
of the external potential $U_{\tbox{ext}} = (4\pi e^2/q^2) \rho_{\tbox{ext}}$, 
and the induced potential $U = (4\pi e^2/q^2) \rho_{\tbox{elct}}$,  
where $\rho_{\tbox{elct}}$ is the total density of the electrons.  
The dielectric constant can be deduced from the equations of 
motion ${\partial \rho_{\tbox{elct}}/ \partial t = -\nabla J}$  
with ${J=-(\sigma/e^2) \nabla U_{\tbox{total}} - D \nabla \rho_{\tbox{elct}}}$ 
that leads to the relation
\beq{98}
\rho_{\tbox{elct}} \ \ = \ \ \frac{(\sigma/e^2) q^2}{i\omega-Dq^2} \ U_{\tbox{total}},
\eeq
and hence to ${U_{\tbox{total}}=(1/\varepsilon)U_{\tbox{ext}}}$, where 
\beq{0}
\varepsilon(q,\omega) = 1 - \frac{4\pi\sigma}{i\omega-Dq^2}.
\eeq
Note that 
\beq{0}
\im\left[ \frac{-1}{\varepsilon(q,\omega)} \right] 
= 
\frac{4\pi\sigma\omega}{(Dq^2+4\pi\sigma)^2 + \omega^2}
\approx
\frac{\omega}{4\pi\sigma}.
\eeq

The interaction between the electrons and an external electrostatic field 
is described by ${\mathcal{H}_{\tbox{ext}} = U_{\tbox{ext}} \rho_{\tbox{elct}}}$ 
which can be also written as ${\mathcal{H}_{\tbox{ext}} = \rho_{\tbox{ext}} U}$.
The fluctuation dissipation relation expresses $\tilde{S}^{\tbox{eq}}(q,\omega)$ 
using the response function  $\alpha(\bmq,\omega)$  
that relates $U$ to $-\rho_{\tbox{ext}}$ which is   
\beq{0}
\alpha(\bmq,\omega) =
\frac{4\pi e^2}{\bmq^2} 
\left[ 1-\frac{1}{\varepsilon(\bmq,\omega)} \right].
\eeq
Namely, 
\beq{100}
\tilde{S}^{\tbox{eq}}(\bmq,\omega) =
\im\Big[\alpha(\bmq,\omega)\Big]
\, \left(\frac{2}{1-\eexp{-\omega/T}}\right),
\eeq
leading to
\beq{103}
\tilde{S}^{\tbox{eq}}(\bmq,\omega) \approx
\frac{e^2}{\sigma}
\frac{1}{\bmq^2}
\left(\frac{2\omega}{1-\eexp{-\omega/T}}\right).
\eeq
The Ohmic behavior is cut-off 
by $|\omega| \lesssim 1/\tau_c$
and $|\bmq|\lesssim {1}/{\ell}$ 
where $\ell=v_{\tbox{F}}\tau_c$ is 
the elastic mean free path, 
and $v_{\tbox{F}}$ is the Fermi velocity.
Recalling the Einstein relation ${\sigma=e^2\nu D}$,
where $\nu=\deltaF^{-1}/L^d$ is the density of states per unit volume, 
we can write this result more conveniently 
as follows: 
\beq{3008}
\tilde{S}^{\tbox{eq}}(\bmq,\omega) \approx
\frac{1}{\nu D\bmq^2}
\left(\frac{2\omega}{1-\eexp{-\omega/T}}\right).
\eeq
Note that the electron charge $e$ cancels out from this final result
for the Nyquist noise spectrum. This well-known fact is due to the
effects of screening: A larger value of the charge would be canceled
by a correspondingly stronger suppression of density fluctuations.



\begin{thebibliography}{99}

\expandafter\ifx\csname natexlab\endcsname\relax\def\natexlab#1{#1}\fi
\expandafter\ifx\csname bibnamefont\endcsname\relax
  \def\bibnamefont#1{#1}\fi
\expandafter\ifx\csname bibfnamefont\endcsname\relax
  \def\bibfnamefont#1{#1}\fi
\expandafter\ifx\csname citenamefont\endcsname\relax
  \def\citenamefont#1{#1}\fi
\expandafter\ifx\csname url\endcsname\relax
  \def\url#1{\texttt{#1}}\fi
\expandafter\ifx\csname urlprefix\endcsname\relax\def\urlprefix{URL }\fi
\providecommand{\bibinfo}[2]{#2}
\providecommand{\eprint}[2][]{\url{#2}}

\bibitem[{\citenamefont{Altshuler et~al.}(1982)\citenamefont{Altshuler, Aronov,
  and Khmelnitskii}}]{AAK}
\bibinfo{author}{\bibfnamefont{B.}~\bibnamefont{Altshuler}},
  \bibinfo{author}{\bibfnamefont{A.}~\bibnamefont{Aronov}}, \bibnamefont{and}
  \bibinfo{author}{\bibfnamefont{D.}~\bibnamefont{Khmelnitskii}},
  \bibinfo{journal}{J. Phys. C} \textbf{\bibinfo{volume}{15}},
  \bibinfo{pages}{7367} (\bibinfo{year}{1982}).

\bibitem[{\citenamefont{Fukuyama and Abrahams}(1983)}]{Fukuyama1983}
\bibinfo{author}{\bibfnamefont{H.}~\bibnamefont{Fukuyama}} \bibnamefont{and}
  \bibinfo{author}{\bibfnamefont{E.}~\bibnamefont{Abrahams}},
  \bibinfo{journal}{Phys. Rev. B} \textbf{\bibinfo{volume}{27}},
  \bibinfo{pages}{5976} (\bibinfo{year}{1983}).

\bibitem[{\citenamefont{Aleiner
  et~al.}(1999{\natexlab{a}})\citenamefont{Aleiner, Altshuler, and
  Gershenson}}]{alt}
\bibinfo{author}{\bibfnamefont{I.~L.} \bibnamefont{Aleiner}},
  \bibinfo{author}{\bibfnamefont{B.~L.} \bibnamefont{Altshuler}},
  \bibnamefont{and} \bibinfo{author}{\bibfnamefont{M.~E.}
  \bibnamefont{Gershenson}}, \bibinfo{journal}{Waves in Random Media}
  \textbf{\bibinfo{volume}{9}}, \bibinfo{pages}{201}
  (\bibinfo{year}{1999}{\natexlab{a}}).

\bibitem[{\citenamefont{Aleiner
  et~al.}(1999{\natexlab{b}})\citenamefont{Aleiner, Altshuler, and
  Gershenson}}]{alt2}
\bibinfo{author}{\bibfnamefont{I.~L.} \bibnamefont{Aleiner}},
  \bibinfo{author}{\bibfnamefont{B.~L.} \bibnamefont{Altshuler}},
  \bibnamefont{and} \bibinfo{author}{\bibfnamefont{M.~E.}
  \bibnamefont{Gershenson}}, \bibinfo{journal}{Phys. Rev. Lett.}
  \textbf{\bibinfo{volume}{82}}, \bibinfo{pages}{3190}
  (\bibinfo{year}{1999}{\natexlab{b}}).

\bibitem[{\citenamefont{Aleiner et~al.}(2002)\citenamefont{Aleiner, Altshuler,
  and Vavilov}}]{Aleiner2002}
\bibinfo{author}{\bibfnamefont{I.}~\bibnamefont{Aleiner}},
  \bibinfo{author}{\bibfnamefont{B.}~\bibnamefont{Altshuler}},
  \bibnamefont{and} \bibinfo{author}{\bibfnamefont{M.}~\bibnamefont{Vavilov}},
  \bibinfo{journal}{arXiv:cond-mat/0208264}  (\bibinfo{year}{2002}).

\bibitem[{\citenamefont{{von Delft} et~al.}(2007)\citenamefont{{von Delft},
  Marquardt, Smith, and Ambegaokar}}]{munich2}
\bibinfo{author}{\bibfnamefont{J.}~\bibnamefont{{von Delft}}},
  \bibinfo{author}{\bibfnamefont{F.}~\bibnamefont{Marquardt}},
  \bibinfo{author}{\bibfnamefont{R.~A.} \bibnamefont{Smith}}, \bibnamefont{and}
  \bibinfo{author}{\bibfnamefont{V.}~\bibnamefont{Ambegaokar}},
  \bibinfo{journal}{Phys. Rev. B} \textbf{\bibinfo{volume}{76}},
  \bibinfo{pages}{195332} (\bibinfo{year}{2007}).

\bibitem[{\citenamefont{Cohen}(1998)}]{dph2}
\bibinfo{author}{\bibfnamefont{D.}~\bibnamefont{Cohen}}, \bibinfo{journal}{J.
  Phys. A} \textbf{\bibinfo{volume}{31}}, \bibinfo{pages}{8199}
  (\bibinfo{year}{1998}).

\bibitem[{\citenamefont{Cohen and Imry}(1999)}]{dph}
\bibinfo{author}{\bibfnamefont{D.}~\bibnamefont{Cohen}} \bibnamefont{and}
  \bibinfo{author}{\bibfnamefont{Y.}~\bibnamefont{Imry}},
  \bibinfo{journal}{Phys. Rev. B} \textbf{\bibinfo{volume}{59}},
  \bibinfo{pages}{11143} (\bibinfo{year}{1999}).

\bibitem[{\citenamefont{Imry}(2002)}]{imry}
\bibinfo{author}{\bibfnamefont{Y.}~\bibnamefont{Imry}},
  \emph{\bibinfo{title}{Introduction to Mesoscopic Physics}}
  (\bibinfo{publisher}{Oxford University Press}, \bibinfo{year}{2002}),
  \bibinfo{edition}{second edition} ed.

\bibitem[{\citenamefont{Marquardt et~al.}(2007)\citenamefont{Marquardt, {von
  Delft}, Smith, and Ambegaokar}}]{munich1}
\bibinfo{author}{\bibfnamefont{F.}~\bibnamefont{Marquardt}},
  \bibinfo{author}{\bibfnamefont{J.}~\bibnamefont{{von Delft}}},
  \bibinfo{author}{\bibfnamefont{R.~A.} \bibnamefont{Smith}}, \bibnamefont{and}
  \bibinfo{author}{\bibfnamefont{V.}~\bibnamefont{Ambegaokar}},
  \bibinfo{journal}{Phys. Rev. B} \textbf{\bibinfo{volume}{76}},
  \bibinfo{pages}{195331} (\bibinfo{year}{2007}).

\bibitem[{\citenamefont{Stern et~al.}(1990)\citenamefont{Stern, Aharonov, and
  Imry}}]{SAI}
\bibinfo{author}{\bibfnamefont{A.}~\bibnamefont{Stern}},
  \bibinfo{author}{\bibfnamefont{Y.}~\bibnamefont{Aharonov}}, \bibnamefont{and}
  \bibinfo{author}{\bibfnamefont{Y.}~\bibnamefont{Imry}},
  \bibinfo{journal}{Phys. Rev. A} \textbf{\bibinfo{volume}{41}},
  \bibinfo{pages}{3436} (\bibinfo{year}{1990}).


\bibitem[{\citenamefont{Cohen and Horovitz}(2007)}]{rgd}
\bibinfo{author}{\bibfnamefont{D.}~\bibnamefont{Cohen}} \bibnamefont{and}
  \bibinfo{author}{\bibfnamefont{B.}~\bibnamefont{Horovitz}},
  \bibinfo{journal}{J. Phys. A} \textbf{\bibinfo{volume}{40}},
  \bibinfo{pages}{12281} (\bibinfo{year}{2007}).

\bibitem[{\citenamefont{Cohen and Horovitz}(2008)}]{rgd2}
\bibinfo{author}{\bibfnamefont{D.}~\bibnamefont{Cohen}} \bibnamefont{and}
  \bibinfo{author}{\bibfnamefont{B.}~\bibnamefont{Horovitz}},
  \bibinfo{journal}{Europhysics Letters} \textbf{\bibinfo{volume}{81}},
  \bibinfo{pages}{30001} (\bibinfo{year}{2008}).


\bibitem[{\citenamefont{Chakravarty and Schmid}(1986)}]{cak}
\bibinfo{author}{\bibfnamefont{S.}~\bibnamefont{Chakravarty}} \bibnamefont{and}
  \bibinfo{author}{\bibfnamefont{A.}~\bibnamefont{Schmid}},
  \bibinfo{journal}{Phys. Rep.} \textbf{\bibinfo{volume}{140}},
  \bibinfo{pages}{193} (\bibinfo{year}{1986}).

\bibitem[{\citenamefont{Golubev and Zaikin}(1998)}]{zaikin}
\bibinfo{author}{\bibfnamefont{D.~S.} \bibnamefont{Golubev}} \bibnamefont{and}
  \bibinfo{author}{\bibfnamefont{A.~D.} \bibnamefont{Zaikin}},
  \bibinfo{journal}{Phys. Rev. Lett.} \textbf{\bibinfo{volume}{81}},
  \bibinfo{pages}{1074} (\bibinfo{year}{1998}).

\bibitem[{\citenamefont{Golubev and Zaikin}(1999)}]{zaikin2}
\bibinfo{author}{\bibfnamefont{D.~S.} \bibnamefont{Golubev}} \bibnamefont{and}
  \bibinfo{author}{\bibfnamefont{A.~D.} \bibnamefont{Zaikin}},
  \bibinfo{journal}{Phys. Rev. B} \textbf{\bibinfo{volume}{59}},
  \bibinfo{pages}{9195} (\bibinfo{year}{1999}).

\bibitem[{\citenamefont{von Delft}(2008)}]{Delft2008}
\bibinfo{author}{\bibfnamefont{J.}~\bibnamefont{von Delft}},
  \bibinfo{journal}{International Journal of Modern Physics B}
  \textbf{\bibinfo{volume}{22}}, \bibinfo{pages}{727} (\bibinfo{year}{2008}),
  \bibinfo{note}{arXiv:cond-mat/0510563}.

\bibitem{asym1}
G.B. Lesovik and R. Loosen, 
JETP Lett. {\bf 65}, 269 (1997).

\bibitem{asym2}
U. Gavish, Y. Levinson, and Y. Imry,
Phys. Rev. B {\bf 62}, R10637 (2000).



\bibitem[{\citenamefont{Clerk et~al.}(accepted, 2009)\citenamefont{Clerk,
  Devoret, Girvin, Marquardt, and
  Schoelkopf}}]{2008_10_ClerkEtAl_QuantumNoiseReview}
\bibinfo{author}{\bibfnamefont{A.~A.} \bibnamefont{Clerk}},
  \bibinfo{author}{\bibfnamefont{M.~H.} \bibnamefont{Devoret}},
  \bibinfo{author}{\bibfnamefont{S.~M.} \bibnamefont{Girvin}},
  \bibinfo{author}{\bibfnamefont{F.}~\bibnamefont{Marquardt}},
  \bibnamefont{and} \bibinfo{author}{\bibfnamefont{R.~J.}
  \bibnamefont{Schoelkopf}}, \bibinfo{journal}{arXiv:0810.4729, Reviews of
  Modern Physics}  (\bibinfo{year}{accepted, 2009}).


\bibitem{weiss}
U. Weiss,
{\em Quantum Dissipative Systems} (World Scientific, Singapore, 1999).



\bibitem[{\citenamefont{Marquardt}(2005)}]{2005_Marquardt_MZQB_EPL}
\bibinfo{author}{\bibfnamefont{F.}~\bibnamefont{Marquardt}},
  \bibinfo{journal}{Europhysics Letters} \textbf{\bibinfo{volume}{72}},
  \bibinfo{pages}{788} (\bibinfo{year}{2005}).

\bibitem[{\citenamefont{Marquardt}(2006)}]{2006_04_MZQB_Long}
\bibinfo{author}{\bibfnamefont{F.}~\bibnamefont{Marquardt}},
  \bibinfo{journal}{Phys. Rev. B} \textbf{\bibinfo{volume}{74}},
  \bibinfo{pages}{125319} (\bibinfo{year}{2006}).

\bibitem[{\citenamefont{Marquardt and Golubev}(2004)}]{2004_Marquardt_DFS_PRL}
\bibinfo{author}{\bibfnamefont{F.}~\bibnamefont{Marquardt}} \bibnamefont{and}
  \bibinfo{author}{\bibfnamefont{D.~S.} \bibnamefont{Golubev}},
  \bibinfo{journal}{Phys. Rev. Lett.} \textbf{\bibinfo{volume}{93}},
  \bibinfo{pages}{130404} (\bibinfo{year}{2004}).

\bibitem[{\citenamefont{Marquardt and
  Golubev}(2005)}]{2004_Marquardt_DFS_LongVersion}
\bibinfo{author}{\bibfnamefont{F.}~\bibnamefont{Marquardt}} \bibnamefont{and}
  \bibinfo{author}{\bibfnamefont{D.~S.} \bibnamefont{Golubev}},
  \bibinfo{journal}{Phys. Rev. A} \textbf{\bibinfo{volume}{72}},
  \bibinfo{pages}{022113} (\bibinfo{year}{2005}).

\bibitem{Ze1}
O. Nairz, M. Arndt and A. Zeilinger,
American Journal of Physics {\bf 71}, 319 (2003).

\bibitem{Ze2}
B. Brezger, L. Hackermüller, S. Uttenthaler, J. Petschinka, M. Arndt, and A. Zeilinger, 
Phys. Rev. Lett. {\bf 88}, 100404 (2002).


\ \\

\bibitem{rmA} For the purpose of this illustration we assume that 
the velocity and hence the diffusion coefficient of the electron  
do not depend on its kinetic energy (a linear dispersion relation). 
The electrons are added one by one, and at low densities 
form a Boltzmann gas. Only when the chemical potential 
exceeds~$T$ one starts to observe the implications of Fermi statistics.      

\bibitem{rmB} To label one or several electrons as test particles 
is of course meaningless if we take their indistinguishablity into account.
So the dashed line has some meaning only in a vague semiclassical context
which we try to bypass. If we literally divided the electron into 
two sub-volumes, the purity loss and hence the many-body dephasing rate 
would be proportional to the area of dividing surface that separate the two spatial regions.
   
\bibitem{rmC}
The notion of ``test particles" acquires a more concrete meaning 
within the context of a transport experiment, where the ``test particles" 
are the ones contributing to the current, or even more to the point, 
in the context to weak localization, to the back-scattered current.


\end{thebibliography}

\clearpage

\end{document}